\documentclass[conference]{IEEEtran}
\IEEEoverridecommandlockouts
\usepackage{cite}
\usepackage{amsmath,amssymb,amsfonts}
\usepackage{algorithmic}
\usepackage{graphicx}
\usepackage{textcomp}
\usepackage{xcolor}
\def\BibTeX{{\rm B\kern-.05em{\sc i\kern-.025em b}\kern-.08em
    T\kern-.1667em\lower.7ex\hbox{E}\kern-.125emX}}

\newtheorem{theorem}{\it Theorem}

\newtheorem{definition}{\it Definition}

\newtheorem{proposition}{\it Proposition}

\begin{document}
	
\title{Two-Way Coding and Attack Decoupling in Control Systems Under Injection Attacks\\
%\title{Two-Way Coding Meets Control: Attack Decoupling in Control Systems Under Injection Attacks\\
\thanks{The work is supported by the Knut and Alice Wallenberg Foundation, the Swedish Strategic Research Foundation, the Swedish Research Council, the Swedish Civil Contingencies Agency (CERCES project), the JSPS under Grant-in-Aid for Scientific Research Grant No.:~15H04020, and the JST CREST under Grant No.:~JPMJCR15K3.}
}

\author{\IEEEauthorblockN{Song Fang, Karl Henrik Johansson, Mikael Skoglund, Henrik Sandberg}
\IEEEauthorblockA{\textit{School of Electrical Engineering and Computer Science} \\
\textit{KTH Royal Institute of Technology} \\
{\tt\small sonf@kth.se, kallej@kth.se, skoglund@kth.se, hsan@kth.se}}
\and
\IEEEauthorblockN{Hideaki Ishii}
\IEEEauthorblockA{\textit{Department of Computer Science} \\
\textit{Tokyo Institute of Technology}\\
{\tt\small ishii@c.titech.ac.jp}}}

\maketitle

\begin{abstract}
In this paper, we introduce the concept of two-way coding, which originates in communication theory characterizing coding schemes for two-way channels, into control theory, particularly to facilitate the analysis and design of feedback control systems under injection attacks. Moreover, we propose the notion of attack decoupling, and show how the controller and the two-way coding can be co-designed to nullify the transfer function from attack to plant, rendering the attack effect zero both in transient phase and in steady state.
\end{abstract}


\section{Introduction}

Observations on the underlying connections between communication and control date back to \cite{blackman1946data}, in which the authors (including Shannon and Bode) stated that ``there is an obvious analogy between the problem of smoothing the data to eliminate or reduce the effect of tracking errors and the problem of separating a signal from interfering noise in communications systems”. In recent years, since the integrations of communication and control systems are becoming more and more prevalent, as witnessed in, e.g., cyber-physical systems and IoT systems, the interaction of communication theory (including information theory and coding theory) and control theory has especially been a heated topic (see, e.g., \cite{fang2017towards} and the references therein). In such interactions, concepts and tools from communication such as entropy have been introduced to control (see, e.g., \cite{Mar:08}), and so are those from control to communication, as in, for instance, \cite{kim2010feedback}.

In this paper, we introduce yet another notion from communication to control: two-way coding in two-way communication. The concept of two-way communication channels was proposed by Shannon \cite{shannon1961two}. As its name indicates, in two-way channels, signals are transmitted simultaneously in both directions between the two terminals of communication. Accordingly, coding schemes for two-way channels should utilize the information contained in the data streaming in both directions. Stated alternatively, the coding schemes should also be two-way, and thus are correspondingly referred to as two-way coding \cite{van1977survey, meeuwissen1998information, chaaban2015multi}.

With the controller side and the plant side being respectively viewed as the two terminals of communication, the communication channels embedded in networked feedback control systems are inherently two-way channels. However, approaches based on two-way coding for the two-way channels in networked feedback systems are rarely seen in the literature. One exception is the so-called scattering transformation utilized in the tele-operation of robotics \cite{hokayem2006bilateral}, although, as far as we know, its connection with two-way coding has never before been established. Nevertheless, scattering transformation can be viewed in a broad sense as a special class of two-way coding, aiming to resolve the issue of two-way time delays, the most essential characterization and the main issue of the two-way channels modeled on the input-output level in the problem of tele-operation. 
%Other related applications of the scattering transformation include \cite{kimura1996chain, kailath2000linear, gu2011two}.

When it comes to cyber-physical security problems arising in networked control systems (see, e.g., \cite{poovendran2012special, johansson2014guest, sandberg2015cyberphysical, teixeira2015secure, zhu2015game, amin2015game, smith2015covert, mo2015physical, pasqualetti2015control, cheng2017guest, giraldo2018survey, chong2019tutorial} and the references therein), to the best of our knowledge, only one-way coding has been employed. The authors of \cite{xu2015secure} introduced one-way encryption matrices into control systems to achieve confidentiality and integrity. In \cite{miao2017coding}, the authors considered using one-way coding matrices to encode the sensor outputs in order to detect stealthy false data injection attacks in cyber-physical systems. 
One-way modulation matrices were inserted into cyber-physical systems in \cite{hoehn2016detection} to detect covert attacks and zero-dynamics attacks. 
Dynamic one-way coding was applied to detect and isolate routing attacks \cite{ferrari2017detection1} and replay attacks \cite{ferrari2017detection2}. 
For remote state estimation in the presence of eavesdroppers, the so-called state-secrecy codes were introduced \cite{tsiamis2017state}, which are also essentially one-way coding schemes.
Nevertheless, one-way coding has its inherent limitations; for instance, one-way coding in general cannot eliminate the unstable poles nor nonminimum-phase zeros of the plant nor the controller \cite{FangICCPS19}, which are most critical issues in the defense against, e.g., zero-dynamics attacks \cite{teixeira2015secure}. 

In our previous work \cite{FangICCPS19}, we examined how the presence of two-way coding in linear time-invariant (LTI) feedback control systems can make the zeros and/or poles of the equivalent plant as viewed by the attacker all different from those of the original plant, and under some additional assumptions (i.e., the plant is stabilizable by static output feedback), the equivalent plant may even be made stable and/or minimum-phase. In the particular case of zero-dynamics attacks, it is then implicated that the attacks will be detected if designed according to the original plant, while the attack effect may be corrected in steady state if the attacks are to be designed with respect to the equivalent plant.

%More specifically, with the aid of two-way coding, the zeros and poles of the ``transformed" plant can be made all different from the original plant; as such, attacks that rely on the knowledge of plant models, e.g., the zero-dynamics attacks, will be detected. 
%%Indeed, the zeros and poles of the ``transformed" controller may also be made all different from the original controller, thus preventing potential attacks that need to know the controller models as well. 
%Additionally, we go one step further and investigate what the consequences of various attacks on the control performance might be; even when after attacks have been detected, knowing the possible attack effect and how to correct it remains to be necessary, especially when the running of the control system cannot be stopped immediately after detecting the attacks. In general, it is seen that for attacks that appear only in the forward path (or only in the feedback path) and with transfer function type $\leq 1$, e.g., the zero-dynamics attacks, the parameters of two-way coding may be such tuned that the attack effect can be eliminated completely in steady state. Meanwhile, when it comes to attacks (with transfer function type $\leq 1$) that are injected in both the forward path and the feedback path simultaneously, e.g., the covert attacks, an ``impossibility theorem" exists and the attack effect is not completely eliminable in steady state.

To prevent possible damages during the transient phase even when the attack affect can be corrected in steady state, in this paper we propose the notion of attack decoupling. For LTI systems, we say that a certain attack is decoupled if the transfer function from attack to plant input/output is made zero, without making zero the transfer function from reference to plant input/output. As such, when attack decoupling is achieved, the attack response will be completely zero both in transient phase and in steady state. We then examine in order conventional feedback systems, feedback systems with one-way coding, as well as feedback systems with two-way coding, and discover that it is only in feedback systems with two-way coding that attacks in the uplink or downlink channels can be decoupled.

The remainder of the paper is organized as follows. Section~II introduces the two-way coding. In Section~III, we propose the notion of attack decoupling. Concluding remarks are given in Section~IV.

\section{Two-Way Coding}

\begin{figure}
	\vspace*{-3mm}
	\begin{center}
		\includegraphics [width=0.5\textwidth]{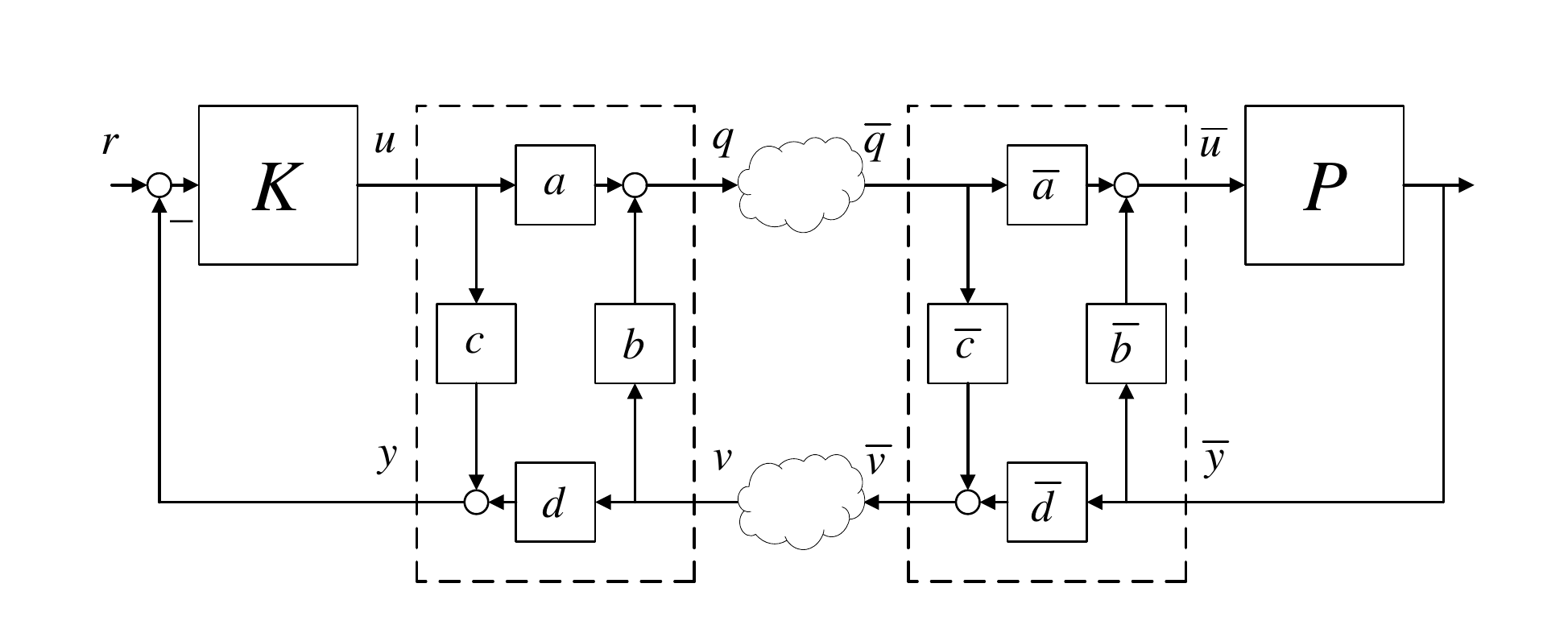}
		\vspace*{-6mm}
		\caption{A networked feedback system with two-way coding.}
		\label{figure1}
	\end{center}
	\vspace*{-3mm}
\end{figure}

Consider the single-input single-output (SISO) system depicted in Fig.~\ref{figure1}. Herein, $K$ denotes the controller while $P$ denotes the plant. The reference signal is $r \left( t \right) \in \mathbb{R}$ and the plant output is $\overline{y} \left( t \right) \in \mathbb{R}$. In addition, let $u \left( t \right)$, $\overline{u} \left( t \right)$, $y \left( t \right)$, $q \left( t \right)$, $\overline{q} \left( t \right)$, $v \left( t \right)$, $\overline{v} \left( t \right) \in \mathbb{R}$. 

\begin{definition}
	The (static) two-way coding is defined as
	\begin{flalign}
	\left[
	\begin{array}{c}
	q \left( t \right)\\
	y \left( t \right)\\
	\end{array}
	\right]
	&=M
	\left[
	\begin{array}{c}
	u \left( t \right)\\
	v \left( t \right)\\
	\end{array}
	\right]
	%&=\left[
	%\begin{array}{cc}
	%a & b \\
	%c & d \\
	%\end{array}\right]
	%\left[
	%\begin{array}{c}
	%U \left( s \right)\\
	%V \left( s \right)\\
	%\end{array}
	%\right]
	,~
	M
	=\left[
	\begin{array}{cc}
	a & b \\
	c & d \\
	\end{array}\right].
	\end{flalign}
	Herein, $a, b, c, d \in \mathbb{R}$ are chosen such that 
	\begin{flalign} \label{condition1}
	ad \neq 0,~ ad - bc \neq 0. 
	\end{flalign}
	Strictly speaking, it should be further assumed that $\left| ad - bc \right| < \infty$.
\end{definition}

\vspace*{0mm}

\begin{figure}
	\vspace*{-3mm}
	\begin{center}
		\includegraphics [width=0.5\textwidth]{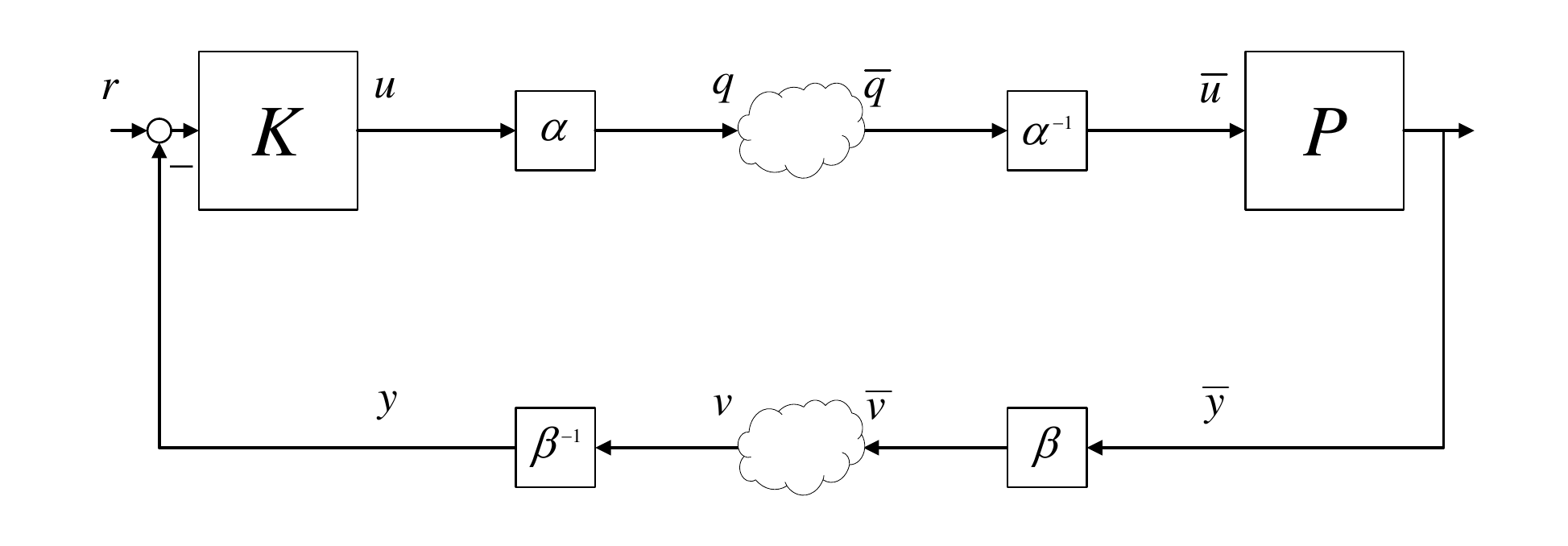}
		\vspace*{-6mm}
		\caption{A networked feedback system with one-way coding.}
		\label{figureoneway}
	\end{center}
	\vspace*{-3mm}
\end{figure}

Herein, two-way coding (that operates in a feedback loop) represents a two-way transformation taking in the signal in the forward path and the signal in the feedback path while outputting a new signal to the forward path and a second new signal that passes on in the feedback path. In comparison, Fig.~\ref{figureoneway} depicts a system with one-way coding schemes, which are one-way transformations that either take in the signal in the forward path and output a new signal that passes on in the forward path, or input the signal in the feedback path and output a signal that continues in the feedback path; herein, $\alpha, \beta \in \mathbb{R}$ and  $0< \left| \alpha \right|, \left| \beta \right| < \infty$.

For simplicity, we denote the inverse of two-way coding $M$ as 
\begin{flalign} \label{inverse}
\left[
\begin{array}{cc}
\overline{a} & \overline{b} \\
\overline{c} & \overline{d} \\
\end{array}\right]
= M^{-1} = \left[
\begin{array}{cc}
\frac{d}{ad-bc} & -\frac{b}{ad-bc} \\
-\frac{c}{ad-bc} & \frac{a}{ad-bc} \\
\end{array}\right],
\end{flalign}
where $\overline{a}, \overline{b}, \overline{c}, \overline{d} \in \mathbb{R}$. As illustrated on the plant side in Fig.~\ref{figure1}, the inverse of two-way coding $M$ denotes another two-way coding.

\subsection{Two-Way Coding in LTI Feedback Control Systems} 

\begin{figure}
	\vspace*{-3mm}
	\begin{center}
		\includegraphics [width=0.5\textwidth]{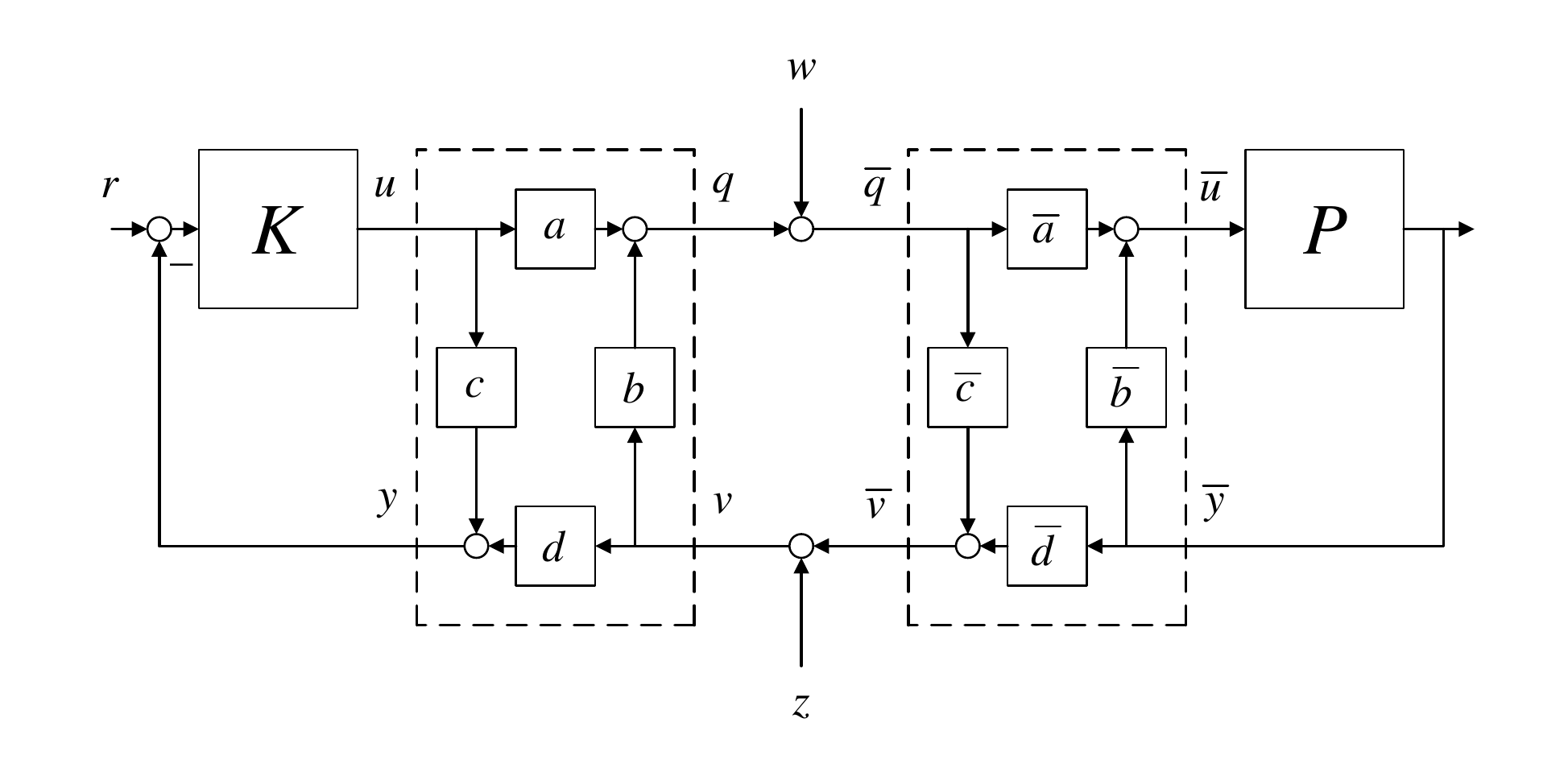}
		\vspace*{-6mm}
		\caption{A feedback system with two-way coding under injection attacks.}
		\label{injection}
	\end{center}
	\vspace*{-3mm}
\end{figure}

We next analyze in particular LTI feedback control systems with two-way coding. Consider the SISO feedback system with two-way coding depicted in Fig.~\ref{injection}. Assume that herein the controller $K$ and plant $P$ are LTI with transfer functions $K \left( s \right)$ and $P \left( s \right)$, respectively. In addition, let $r \left( t \right)$, $u \left( t \right)$, $\overline{u} \left( t \right)$, $y \left( t \right)$, $\overline{y} \left( t \right)$, $q \left( t \right)$, $ \overline{q} \left( t \right)$, $v \left( t \right)$, $\overline{v} \left( t \right) \in \mathbb{R}$. Meanwhile, suppose that injection (additive) attacks $w \left( t \right) \in \mathbb{R}$ and $z \left( t \right) \in \mathbb{R}$ exist in the forward path and feedback path of the control systems, respectively. Let $R \left( s \right)$, $U \left( s \right)$, $ \overline{U} \left( s \right)$, $Y \left( s \right)$, $\overline{Y} \left( s \right)$, $Q \left( s \right)$, $\overline{Q} \left( s \right)$, $V \left( s \right)$, $\overline{V} \left( s \right)$, $W \left( s \right)$, $Z \left( s \right)$ represent the Laplace transforms, assuming that they exist, of the signals $r \left( t \right)$, $u \left( t \right)$, $\overline{u} \left( t \right)$, $y \left( t \right)$, $\overline{y} \left( t \right)$, $q \left( t \right)$, $\overline{q} \left( t \right)$, $v \left( t \right)$, $\overline{v} \left( t \right)$, $ w \left( t \right)$, $z \left( t \right)$. From now on, we assume that all the transfer functions of the systems are with zero initial conditions, unless otherwise specified.

We now provide expressions \cite{FangICCPS19} for the Laplace transforms of the plant input $\overline{u} \left( t \right)$ and the plant output $\overline{y} \left( t \right)$, given reference $r \left( t \right)$ and under injection attacks $w \left( t \right)$ and $z \left( t \right)$.

\begin{proposition} \label{foundation}
	Consider the SISO feedback system with two-way coding under injection attacks depicted in Fig.~\ref{injection}.
	Assume that controller $K$ and plant $P$ are LTI with transfer functions $K \left( s \right)$ and $P \left( s \right)$, respectively, and that the closed-loop system is stable.
	Then, 
	\begin{flalign} \label{injection1}
	\overline{U} \left( s \right)
	&= \frac{K \left( s \right)}{1+ K \left( s \right) P \left( s \right)} R \left( s \right)  + \frac{ a^{-1}\left[ 1 + c K \left( s \right) \right] }{1+ K \left( s \right) P \left( s \right)} W \left( s \right)
	\nonumber \\
	&\ \ \ \  + \frac{ a^{-1} \left[ b - \left( ad - bc \right) K \left( s \right) \right] P \left( s \right)}{1+ K \left( s \right) P \left( s \right)} Z \left( s \right),
	\end{flalign}
	and
	\begin{flalign} \label{injection2}
	\overline{Y} \left( s \right)
	&= \frac{K \left( s \right) P \left( s \right)}{1+ K \left( s \right) P \left( s \right)} R \left( s \right)  + \frac{ a^{-1}\left[ 1 + c K \left( s \right) \right] P \left( s \right)}{1+ K \left( s \right) P \left( s \right)} W \left( s \right)
	\nonumber \\
	&\ \ \ \  + \frac{ a^{-1} \left[ b - \left( ad - bc \right) K \left( s \right) \right] P \left( s \right)}{1+ K \left( s \right) P \left( s \right)} Z \left( s \right).
	\end{flalign}
\end{proposition}

\vspace*{0mm}

Proposition~\ref{foundation} lays the foundation for the analysis of attack decoupling in feedback systems with two-way coding, as will be discussed shortly.

\section{Attack Decoupling} \label{null}

In what follows, we propose the notion of attack decoupling, which features a strong notion of security in the context of cyber-physical systems; in general, however, it is a more broad control-theoretic notion applicable to any (networked) feedback systems. 
%We now present the definition of attack decoupling as follows.

\begin{definition} \label{def1}
	Consider a feedback control system. An attack is said to be decoupled if the attack response in plant
	input/output can be made completely zero for arbitrary attack signals, without nullifying the reference response in plant
	input/output.
\end{definition}

For LTI systems, attack decoupling can be defined more specifically in terms of transfer functions.

\begin{definition} \label{def2}
	Consider an LTI feedback control system. An attack is said to be decoupled if the transfer function from attack to plant input/output can be made zero, without nullifying the transfer function from reference to plant input/output.
\end{definition} 

When the attack is decoupled for a certain attack point, it is as if the path from the attack signal to plant input/output signal is cut off, while not cutting off the signal path from the reference to plant input/output.
In general, attack decoupling is a system-theoretic notion, which is not restricted to dealing with attacks and is more broadly applicable to disturbances and noises. While within the scope of attack analysis, attack decoupling is a strong notion of security, meaning that the attack response will be completely zero both in transient phase and in steady state for arbitrary injection attacks, regardless of what the attacker knows or does.

As a matter of fact, attack decoupling is closely related to the notion of disturbance decoupling in geometric control \cite{wonhambook}. More specifically, disturbance decoupling only requires that the transfer function from the disturbance to plant output to be zero, without requiring the transfer function from the disturbance to plant input to be zero. In this sense, attack decoupling implies and provides a new approach to achieve disturbance decoupling, while bringing new perspectives to other relevant topics in geometric control as well.

We next investigate, one by one, conventional feedback systems, feedback systems with one-way coding, as well as feedback systems with two-way coding, to see whether attack decoupling is possible, and if so, how to achieve it.

\subsection{Conventional Feedback Systems}

\begin{figure}
	\vspace*{-3mm}
	\begin{center}
		\includegraphics[width=0.25\textwidth]{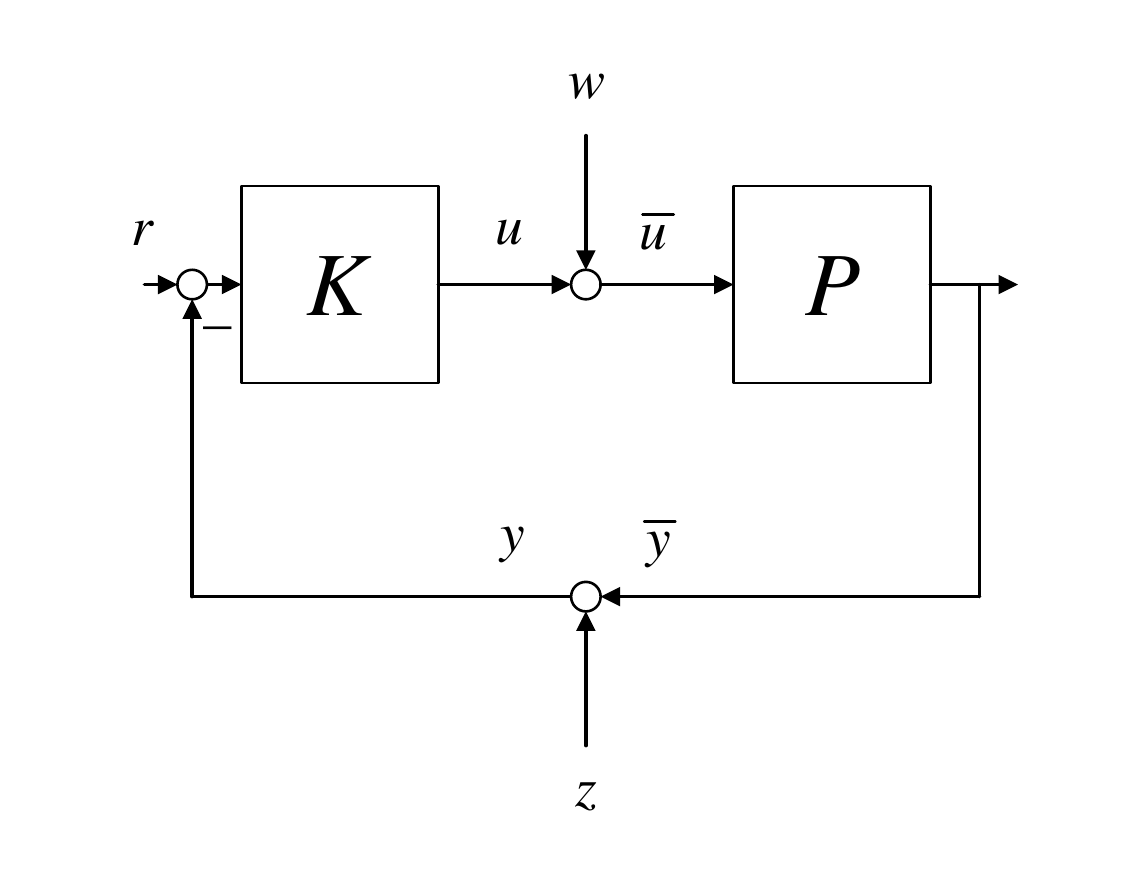}
		\vspace*{-6mm}
		\caption{A feedback system without coding.}
		\label{nullable1}
	\end{center}
	\vspace*{-3mm}
\end{figure}

In the sequel, we let $T_{\overline{u} r} \left(s\right), T_{\overline{u} w} \left(s\right), T_{\overline{u} z} \left(s\right)$ denote the transfer functions from $R \left(s\right), W \left(s\right), Z \left(s\right)$ to $\overline{U} \left(s\right) $, respectively, and let $T_{\overline{y} r} \left(s\right), T_{\overline{y} w} \left(s\right), T_{\overline{y} z} \left(s\right)$ denote the transfer functions from $R \left(s\right), W \left(s\right), Z \left(s\right)$ to $\overline{Y} \left(s\right) $, respectively. We shall first show that or conventional feedback control systems without coding, as depicted in Fig.~\ref{nullable1}, neither attack $w \left(t\right)$ nor attack $z \left(t\right)$ can be decoupled. 

\begin{theorem} \label{nocoding}
	Consider the SISO feedback system depicted in Fig.~\ref{nullable1}. Then, neither attack $w \left(t\right)$ nor attack $z \left(t\right)$ can be decoupled.
\end{theorem}

%\frac{K \left(s\right)}{1 + K \left(s\right) P \left(s\right)} R \left(s\right) 
%+　\frac{1}{1 + K \left(s\right) P \left(s\right)} W \left(s\right)
%- \frac{K \left(s\right)}{1 + K \left(s\right) P \left(s\right)} Z \left(s\right)

\begin{IEEEproof}
	For the system in Fig.~\ref{nullable1}, it can be obtained that
	\begin{flalign}
	\overline{U} \left( s \right) 
	&=  \frac{K \left(s\right)}{1 + K \left(s\right) P \left(s\right)} R \left(s\right)
	+ \frac{1}{1 + K \left(s\right) P \left(s\right)} W \left(s\right) \nonumber\\
	&\ \ \ \ - \frac{K \left(s\right)}{1 + K \left(s\right) P \left(s\right)} Z \left(s\right)
	\nonumber \\ 
	&= T_{\overline{u} r} \left(s\right) R \left(s\right) 
	+ T_{\overline{u} w} \left(s\right) W \left(s\right) 
	+ T_{\overline{u} z} \left(s\right) Z \left(s\right). \nonumber
	\end{flalign}  
	Clearly, 
	\begin{flalign}
	T_{\overline{u}r} \left(s\right) = K \left(s\right) T_{\overline{u}w} \left(s\right)= - T_{\overline{u}z} \left(s\right). \nonumber
	\end{flalign}
	As such, if $T_{\overline{u}w} \left(s\right) = 0$ or $T_{\overline{u}z} \left(s\right) = 0$, then $T_{\overline{u}r} \left(s\right) = 0$. In other words, attack $w \left(t\right)$ cannot be decoupled. Similarly, by noting that
	\begin{flalign}
	\overline{Y} \left( s \right) 
	&=  \frac{K \left(s\right) P \left(s\right)}{1 + K \left(s\right) P \left(s\right)} R \left(s\right)
	+ \frac{P \left(s\right)}{1 + K \left(s\right) P \left(s\right)} W \left(s\right) \nonumber\\
	&\ \ \ \ - \frac{K \left(s\right)P \left(s\right)}{1 + K \left(s\right) P \left(s\right)} Z \left(s\right)
	\nonumber \\ 
	&= T_{\overline{y} r} \left(s\right) R \left(s\right) 
	+ T_{\overline{y} w} \left(s\right) W \left(s\right) 
	+ T_{\overline{y} z} \left(s\right) Z \left(s\right), \nonumber
	\end{flalign}
	and that 
	\begin{flalign}
	T_{\overline{y}r} \left(s\right) = K \left(s\right) T_{\overline{y}w} \left(s\right)= - T_{\overline{y}z} \left(s\right), \nonumber
	\end{flalign}
	it follows that attack $z \left(t\right)$ cannot be decoupled.
\end{IEEEproof}

\subsection{Feedback System with One-Way Coding}

\begin{figure}
	\vspace*{-3mm}
	\begin{center}
		\includegraphics[width=0.5\textwidth]{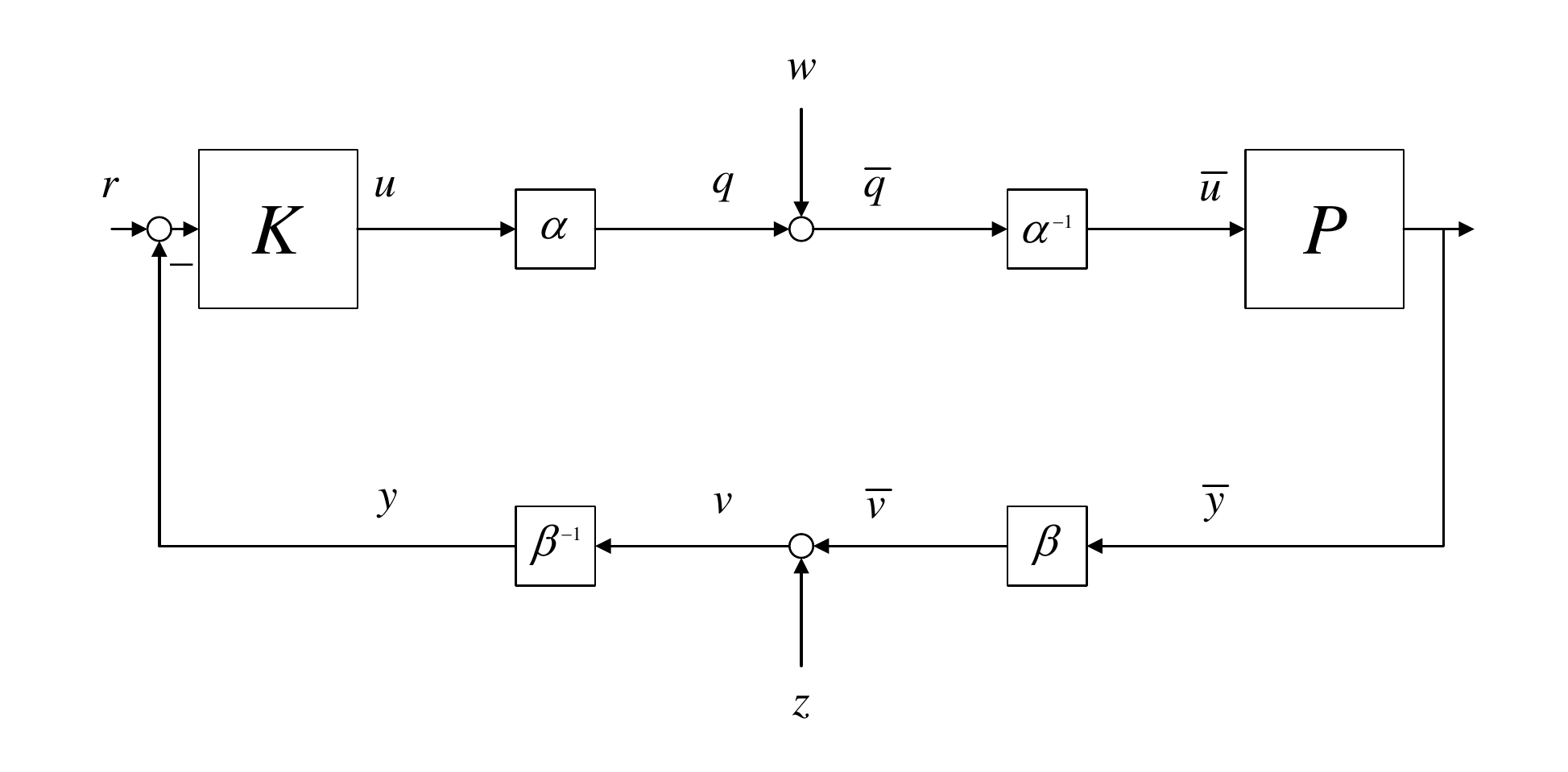}
		\vspace*{-6mm}
		\caption{A feedback system with one-way coding.}
		\label{nullable2}
	\end{center}
	\vspace*{-3mm}
\end{figure}

In the system in Fig.~\ref{nullable2} with one-way coding, neither attack $w \left(t\right)$ nor attack $z \left(t\right)$ can be decoupled.

\begin{theorem} \label{onewaycoding}
	Consider the SISO feedback system depicted in Fig.~\ref{nullable2}. Then, neither attack $w \left(t\right)$ nor attack $z \left(t\right)$ can be decoupled.
\end{theorem}

%\frac{K \left(s\right)}{1 + K \left(s\right) P \left(s\right)} R \left(s\right) 
%+　\frac{1}{1 + K \left(s\right) P \left(s\right)} W \left(s\right)
%- \frac{K \left(s\right)}{1 + K \left(s\right) P \left(s\right)} Z \left(s\right)

\begin{IEEEproof}
	For the system in Fig.~\ref{nullable2}, it can be obtained that
	\begin{flalign}
	\overline{U} \left( s \right) 
	&=  \frac{K \left(s\right)}{1 + K \left(s\right) P \left(s\right)} R \left(s\right)
	+ \frac{\alpha^{-1}}{1 + K \left(s\right) P \left(s\right)} W \left(s\right) \nonumber\\
	&\ \ \ \ - \frac{\beta^{-1}K \left(s\right)}{1 + K \left(s\right) P \left(s\right)} Z \left(s\right)
	\nonumber \\ 
	&= T_{\overline{u} r} \left(s\right) R \left(s\right) 
	+ T_{\overline{u} w} \left(s\right) W \left(s\right) 
	+ T_{\overline{u} z} \left(s\right) Z \left(s\right). \nonumber
	\end{flalign}
	Clearly, 
	\begin{flalign}
	T_{\overline{u}r} \left(s\right) = \alpha K \left(s\right) T_{\overline{u}w} \left(s\right)= - \beta T_{\overline{u}z} \left(s\right). \nonumber
	\end{flalign}
	As such, if $T_{\overline{u}w} \left(s\right) = 0$ or $T_{\overline{u}z} \left(s\right) = 0$, then $T_{\overline{u}r} \left(s\right) = 0$. In other words, attack $w \left(t\right)$ cannot be decoupled. Similarly, by noting that
	\begin{flalign}
	\overline{Y} \left( s \right) 
	&=  \frac{K \left(s\right) P \left(s\right)}{1 + K \left(s\right) P \left(s\right)} R \left(s\right)
	+ \frac{\alpha^{-1} P \left(s\right)}{1 + K \left(s\right) P \left(s\right)} W \left(s\right) \nonumber\\
	&\ \ \ \ - \frac{\beta^{-1} K \left(s\right)P \left(s\right)}{1 + K \left(s\right) P \left(s\right)} Z \left(s\right)
	\nonumber \\ 
	&= T_{\overline{y} r} \left(s\right) R \left(s\right) 
	+ T_{\overline{y} w} \left(s\right) W \left(s\right) 
	+ T_{\overline{y} z} \left(s\right) Z \left(s\right), \nonumber
	\end{flalign}
	and that
	\begin{flalign}
	T_{\overline{y}r} \left(s\right) = \alpha K \left(s\right) T_{\overline{y}w} \left(s\right)= - \beta T_{\overline{y}z} \left(s\right), \nonumber
	\end{flalign}
	it follows that attack $z \left(t\right)$ cannot be decoupled.
\end{IEEEproof}

In fact, it can be shown that attack decoupling on neither points is possible even with dynamic one-way coding $\alpha \left(s\right)$ and $\beta \left(s\right)$.

\subsection{Feedback System with Two-Way Coding}

%\begin{figure}
%	\vspace*{-0mm}
%	\begin{center}
%		\includegraphics[width=0.5\textwidth]{figure1.pdf}
%		\vspace*{-9mm}
%		\caption{A feedback system with two-way coding.}
%		\label{nullable3}
%	\end{center}
%	\vspace*{-6mm}
%\end{figure}

For the system shown in Fig.~\ref{injection} with two-way coding, attack $w \left(t\right)$ can be decoupled, and attack $z \left(t\right)$ can be decoupled as well.

\begin{theorem} \label{nullable}
	Consider the SISO feedback system depicted in Fig.~\ref{injection}. Suppose that plant $P \left(s\right)$ is stabilizable by static output feedback, and that controller $K \left(s\right)$ is chosen among such static output-feedback stabilizing controllers, i.e., $K \left(s\right) = K \in \mathbb{R}$. 
	\begin{itemize}
		\item	If $c= -1/K$, then attack $w \left(t\right)$ is decoupled;
		\item	If $b= \left(ad-bc\right) K$, then attack $z \left(t\right)$ is decoupled.
	\end{itemize}
\end{theorem}

%\frac{K \left(s\right)}{1 + K \left(s\right) P \left(s\right)} R \left(s\right) 
%+　\frac{1}{1 + K \left(s\right) P \left(s\right)} W \left(s\right)
%- \frac{K \left(s\right)}{1 + K \left(s\right) P \left(s\right)} Z \left(s\right)

\begin{IEEEproof}
	%	Denote the sub-systems from $w \left(t\right)$ and $z \left(t\right)$ to plant input $\overline{u} \left(t\right)$ by $T_{\overline{u}w}$ and $T_{\overline{u}z}$, respectively, and similarly, denote the sub-systems from $w \left(t\right)$ and $z \left(t\right)$ to plant output $\overline{y} \left(t\right)$ by $T_{\overline{y}w}$ and $T_{\overline{y}z}$, respectively.
	For the system in Fig.~\ref{injection}, it can be obtained that (see Proposition~\ref{foundation})
	\begin{flalign}
	\overline{U} \left( s \right) 
	&=  \frac{K \left(s\right)}{1 + K \left(s\right) P \left(s\right)} R \left(s\right)
	+ \frac{a^{-1} \left[ 1 + c K \left(s\right) \right]}{1 + K \left(s\right) P \left(s\right)} W \left(s\right) \nonumber\\
	&\ \ \ \ - \frac{a^{-1} \left[ b - \left( ad-bc \right) K \left(s\right) \right]}{1 + K \left(s\right) P \left(s\right)} Z \left(s\right)
	\nonumber \\ 
	&= T_{\overline{u} r} \left(s\right) R \left(s\right) 
	+ T_{\overline{u} w} \left(s\right) W \left(s\right) 
	+ T_{\overline{u} z} \left(s\right) Z \left(s\right), \nonumber
	\end{flalign}
	and 
	\begin{flalign}
	\overline{Y} \left( s \right) 
	&=  \frac{K \left(s\right) P \left(s\right)}{1 + K \left(s\right) P \left(s\right)} R \left(s\right) \nonumber\\
	&\ \ \ \ + \frac{a^{-1} \left[ 1 + c K \left(s\right) \right] P \left(s\right)}{1 + K \left(s\right) P \left(s\right)} W \left(s\right) \nonumber\\
	&\ \ \ \ - \frac{a^{-1} \left[ b - \left( ad-bc \right) K \left(s\right) \right] P \left(s\right)}{1 + K \left(s\right) P \left(s\right)} Z \left(s\right)
	\nonumber \\ 
	&= T_{\overline{y} r} \left(s\right) R \left(s\right) 
	+ T_{\overline{y} w} \left(s\right) W \left(s\right) 
	+ T_{\overline{y} z} \left(s\right) Z \left(s\right), \nonumber
	\end{flalign}
	Clearly, when $K \left(s\right) = K$ and $c= -1/K$, we have 
	\begin{flalign} 
	1 + c K \left(s\right) = 1 + c K = 0, \nonumber
	\end{flalign}
	and it follows that $T_{\overline{u}w} \left(s\right) = 0$ and $T_{\overline{y}w} \left(s\right) = 0$, while $T_{\overline{u}r} \left(s\right) \neq 0$ and $T_{\overline{y}r} \left(s\right) \neq 0$. In other words, attack $w \left(t\right)$ is decoupled. Similarly, when $K \left(s\right) = K$ and $b = \left( ad-bc \right) K$, we have 
	\begin{flalign} 
	b - \left( ad-bc \right) K \left(s\right) = b - \left( ad-bc \right) K = 0, \nonumber
	\end{flalign}
	and it follows that $T_{\overline{u}z} \left(s\right) = 0$ and $T_{\overline{y}z} \left(s\right) = 0$, while $T_{\overline{u}r} \left(s\right) \neq 0$ and $T_{\overline{y}r} \left(s\right) \neq 0$. In other words, attack $z \left(t\right)$ is decoupled. 
\end{IEEEproof}

Intuitively, in feedback systems without coding as well as systems with one-way coding, there is only one feedback loop; as such, if the path from the attack signal to plant input/output signal is to be cut off, then the signal path from the reference to plant input/output will inevitably also be cut off. On the other hand, the presence of two-way coding brings additional feedback loops into a feedback system, enabling, probably in a subtle way, the cutting off of the path from the attack signal to plant input/output signal without cutting off that from the the reference to plant input/output.

Note that the attack decoupling of $w \left(t\right)$ or $z \left(t\right)$ requires the co-design of the controller and the two-way coding, as well as the sacrifice of control performance since controllers are limited to be static output-feedback.

Note also that the conditions for achieving attack decoupling do not involve the plant. In other words, attack decoupling will not be affected by the inaccuracies/uncertainties in plant model $P \left(s\right) $.

It is clear that if attack $w \left(t\right)$ (or $z \left(t\right)$) is decoupled, then its attack response (for any injection attacks) will be zero in plant input/output, even though the attacks may not be detected; for instance, zero-dynamics attacks \cite{teixeira2015secure} at $w \left(t\right)$ (or $z \left(t\right)$) cannot be detected if designed properly, but they will have no influence on plant input/output when attack $w \left(t\right)$ (or $z \left(t\right)$) is decoupled.

\subsubsection{Control-theoretic implications of attack decoupling}
We now examine the implications of attack decoupling in control systems. It is clear that when $T_{\overline{u}w} \left(s\right) = 0$, its $H_{\infty}$ norm is zero, that is,
	\begin{flalign} \label{almost}
	\left\| T_{\overline{u}w} \left(s\right) \right\|_{\infty} = 0,
	\end{flalign}
	and moreover, the corresponding Bode integral is given by
	\begin{flalign} \label{almost2}
	\frac{1}{2 \pi} \int_{-\infty}^{\infty} \ln \left| T_{\overline{u}w} \left( \mathrm{j} \omega \right) \right| \mathrm{d} \omega  = - \infty.
	\end{flalign}
	This means perfect attack attenuation, and thus no limitations (e.g., lower bound on $H_{\infty}$ norm \cite{seron2012fundamental}) or trade-offs (e.g., Bode integral \cite{seron2012fundamental}) are present. Similar conclusions hold for the cases when $T_{\overline{y}w} \left(s\right) = 0$, $T_{\overline{u}z} \left(s\right) = 0$, and $T_{\overline{y}z} \left(s\right) = 0$ as well.

	\subsubsection{Almost vs. exact attack decoupling}
	
	Strictly and practically speaking, for $w \left(t\right)$, only ``almost attack
	decoupling" is possible; otherwise, algebraic loops will be present in the feedback system. 
	By ``almost attack decoupling", we mean the attack signal can be attenuated to any degree of accuracy, e.g., the $H_{\infty}$ norm of the corresponding transfer function can be made arbitrarily close to zero; cf. almost disturbance decoupling in \cite{weiland1989almost}. In particular, \eqref{almost} will then become
	\begin{flalign}
	\left\| T_{\overline{u}w} \left(s\right) \right\|_{\infty} = \epsilon,~\forall \epsilon > 0,
	\end{flalign}
	and \eqref{almost2} becomes
	\begin{flalign}
	\frac{1}{2 \pi} \int_{-\infty}^{\infty} \ln \left| T_{\overline{u}w} \left( \mathrm{j} \omega \right) \right| \mathrm{d} \omega  = \ln \varepsilon,~\forall \varepsilon > 0.
	\end{flalign}
	Meanwhile, for $z \left(t\right)$, ``exact attack decoupling" (as defined in Definition~\ref{def1} and Definition~\ref{def2}) can be achieved. That is to say, the following are still achievable:
	\begin{flalign}
	\left\| T_{\overline{u}z} \left(s\right) \right\|_{\infty} = 0,
	\end{flalign}
	and
	\begin{flalign}
	\frac{1}{2 \pi} \int_{-\infty}^{\infty} \ln \left| T_{\overline{u}z} \left( \mathrm{j} \omega \right) \right| \mathrm{d} \omega  = - \infty.
	\end{flalign}
	However, since almost attack decoupling will involve dynamic $K \left( s \right)$ and/or
	dynamic
	$$
	\left[
	\begin{array}{cc}
	a \left(s\right) & b \left(s\right) \\
	c \left(s\right) & d \left(s\right) \\
	\end{array}\right],$$ 
	we leave the detailed discussions on this topic to future research.

\subsection{An ``Impossibility Theorem"}

In what has been presented in this paper so far, we only considered the so-called ``single-point" attacks, such as zero-dynamics attacks \cite{teixeira2015secure}, where attacks are injected into the feedback system at only one point; in other words, one of $w \left(t\right)$ and $z \left(t\right)$ is zero. In fact, ``double-point" attacks, such as covert attacks \cite{smith2015covert}, where attacks are injected into the system at two points simultaneously, have also been discussed in the literature. 

In the subsequent theorem, it will be shown that for double-point attacks injecting attack signals $w \left(t\right)$ and $z \left(t\right)$ at the same time (e.g., covert attacks \cite{smith2015covert}), the attacks $w \left(t\right)$ and $z \left(t\right)$ cannot be decoupled simultaneously, and hence the attack effect cannot be made completely zero for arbitrary double-point attacks.

\begin{theorem} \label{impossible}
	In the SISO feedback system depicted in Fig.~\ref{injection}, attack $w \left(t\right)$ and attack $z \left(t\right)$ cannot be decoupled simultaneously.
\end{theorem}

\begin{IEEEproof}
	If plant $P \left(s\right)$ is not stabilizable by static output feedback, then neither attack $w \left(t\right)$ nor attack $z \left(t\right)$ can be decoupled. Otherwise, if plant $P \left(s\right)$ is stabilizable by static output feedback, it is known from Theorem~\ref{nullable} that the decoupling of $w \left(t\right)$ requires that $c= -1/K$, while the decoupling of $z \left(t\right)$ requires that $b= \left(ad-bc\right) K$, in addition to that controller $K \left(s\right)$ is chosen among static output-feedback stabilizing controllers, i.e., $K \left(s\right) = K \in \mathbb{R}$, in both cases. That is to say, if $w \left(t\right)$ and $z \left(t\right)$ are to be decoupled simultaneously, then $b= \left(ad-bc\right) K = ad K - bc K = ad K + b$, which leads to $adK = 0$ and contradicts the fact $ad \neq 0$ and $K \neq 0$ (otherwise the transfer function from reference $R \left(s\right)$ to plant input/output, see the proof of Theorem~\ref{nullable}, will be zero).
\end{IEEEproof}

As a matter of fact, it can be shown more generally that even with dynamic two-way coding 
$$
\left[
\begin{array}{cc}
a \left(s\right) & b \left(s\right) \\
c \left(s\right) & d \left(s\right) \\
\end{array}\right],$$ 
simultaneous decoupling of attack $w \left(t\right)$ and attack $z \left(t\right)$ is not possible.

This ``impossibility theorem" characterizes on a fundamental level why double-point attacks are in general more difficult to defend against than single-point attacks. 
We will, however, leave the discussions on the defense against such double-point attacks to future research.

One final remark would be that some of the previous results and discussions apply as well to disturbance decoupling. In particular, it is known from Theorem~\ref{nocoding} and Theorem~\ref{onewaycoding} that therein disturbance decoupling is not possible without coding or with one-way coding, but it can be achieved with two-way coding as shown in Theorem~\ref{nullable}. Additionally, it follows from Theorem~\ref{impossible} that therein disturbance decoupling cannot be achieved in the uplink and downlink channels simultaneously even with two-way coding.

%\subsection{Some Remarks on the Interaction of Communication and Control}
%
%We started this paper by introducing the concept of two-way coding from communication to control. Now let us conclude the paper with several questions that might take all this back to communication: What are the implications of attack decoupling in the context of communication systems (with feedback)? In other words, what are the consequences when the attack signal is decoupled at a certain point of a communication system? What about disturbance decoupling? Will such notions bring new perspectives to communication system design, concerning, e.g., error correction?

\section{Conclusions}

In this paper, we have introduced the method of two-way coding from communication into control, in particular, feedback control systems under injection attacks. 
%We have shown that the presence of two-way coding can distort the perspective of the attacker on the control system; this distorted view on the attacker side was demonstrated to facilitate detecting the attacks, or restricting what the attacker can do, or even correcting the attack effect in steady state. 
Additionally, we have proposed the notion of attack decoupling, and it was seen that the controller and two-way coding can be co-designed to nullify the transfer function from attack to plant, zeroing the attack effect completely both in transient phase and in steady state.
Future research directions include the analysis of dynamic two-way coding, MIMO systems, discrete-time systems, as well as other classes of attacks in the presence of two-way coding. We are also interested in examining the implications of attack decoupling in communication system design, concerning, e.g., error correction. 

%This paragraph will end the body of this sample document.
%Remember that you might still have Acknowledgments or
%Appendices; brief samples of these
%follow.  There is still the Bibliography to deal with; and
%we will make a disclaimer about that here: with the exception
%of the reference to the \LaTeX\ book, the citations in
%this paper are to articles which have nothing to
%do with the present subject and are used as
%examples only.

\bibliographystyle{IEEEtran}
\bibliography{sample-bibliography}

\end{document}